\documentclass{webofc}
\usepackage[varg]{txfonts}   
\usepackage{cleveref}
\usepackage{xspace}
\usepackage[cachedir=./,frozencache=true]{minted}
\setminted{fontsize=\small,linenos,numberblanklines=false,framesep=1mm}
\setminted{frame=none}
\usepackage{xcolor}
\usepackage{wrapfig}
\usepackage{siunitx}
\newcommand{\RooFit}{\texttt{RooFit}\xspace}
\newcommand{\RooStats}{\texttt{RooStats}\xspace}
\newcommand{\HistFactory}{\texttt{HistFactory}\xspace}
\newcommand{\ROOT}{\texttt{ROOT}\xspace}
\newcommand{\PDF}{\texttt{PDF}\xspace}
\newcommand{\PDFs}{\texttt{PDF}s\xspace}

\newcommand{\eg}{e.g.\xspace}
\begin{document}
\title{A Faster, More Intuitive RooFit}
%
%

\author{\firstname{Stephan} \lastname{Hageb\"ock}\inst{1}\fnsep\thanks{\email{stephan.hageboeck@cern.ch}}
}

\institute{CERN, Esplanade des Particules 1, 1211 Geneva 23, Switzerland}

\abstract{%
\RooFit and \RooStats, the toolkits for statistical modelling in \ROOT, are used in most searches and measurements at the Large Hadron Collider as well as at $B$ factories. Larger datasets to be collected at \eg\ the High-Luminosity LHC will enable measurements with higher precision, but will require faster data processing to keep fitting times stable.
In this work, a simplification of \RooFit's interfaces and a redesign of its internal dataflow is presented. Interfaces are being extended to look and feel more STL-like to be more accessible both from C++ and Python to improve interoperability and ease of use, while maintaining compatibility with old code. The redesign of the dataflow improves cache locality and data loading, and can be used to process batches of data with vectorised \texttt{SIMD} computations. This reduces the time for computing unbinned likelihoods by a factor four to 16. This will allow to fit larger datasets of the future in the same time or faster than today’s fits.
}
\maketitle
\section{Introduction}
\RooFit~\cite{RooFit} is a C++ package for statistical modelling distributed with \ROOT~\cite{ROOT}.
\RooFit allows to define computation graphs to connect observables, parameters, functions and \PDFs \footnote{\texttt{P}robability \texttt{D}ensity \texttt{F}unctions} to likelihood models, which can be fit to data and be used for statistical tests.
\RooFit is shipped with \RooStats, a toolkit for performing statistical tests with \RooFit
models. It provides tools such as Toy Monte Carlo tests, setting limits and computing significances. Further, \HistFactory provides tools to create \RooFit models from a collection of \ROOT histograms.

\RooFit was originally developed for the BaBar collaboration, and is now used for statistical inference across many experiments in High-Energy Physics, \eg, at $B$ factories and at the Large Hadron Collider. It is crucial for the final steps of most analyses. It was designed for
single-core processors, and was neither optimised for large caches nor \texttt{SIMD} computations. This work stands at the beginning of efforts to
modernise \RooFit to speed up fits, and make it easier to use from both C++ and Python. This will enable researchers to analyse larger datasets and devise more elaborate statistical models.

\section{Modernising \RooFit's Interfaces\label{sec:Interfaces}}
In \RooFit, any collection of mathematical entities such as parameters, observables, functions or \PDFs are saved or passed to functions using the classes \texttt{RooArgSet} and \texttt{RooArgList}, sets and lists of \RooFit objects.

The most common operation is iterating through these collections, both during fitting and when \eg\  inspecting the values of parameters on the user side. This favours array-like data structures like \eg\ \texttt{std::vector}, but internally, \RooFit's collections were using a linked list with optional hash lookup. Iterating through a \RooFit collection requires the following C++ code in \ROOT 6.16:
\begin{minted}{c++}
TIterator* it = pdf.getParameters(obs)->createIterator();
RooAbsArg* p;
while ((p=(RooAbsArg*)it->Next())) {
  p->Print();
}
delete it;
\end{minted}
To speed up iterating and to provide an STL-like interface for \RooFit's collections, the linked list in \RooFit's collections was replaced by a \texttt{std::vector}. Functions such as \mintinline{c++}{begin()}, \mintinline{c++}{end(), size(), operator[]} were implemented to allow for STL-like handling of the collections. In \ROOT 6.18, iterating through the list of observables of a \PDF as above can therefore be achieved as follows:
\begin{minted}{c++}
for (auto p : *pdf.getParameters(obs))
  p->Print();
\end{minted}
The STL-like interface allows to
reduce heap allocations, and replaces \mintinline{c++}{while (p = Next())} loops by range-based for loops. This reduces code clutter, memory leaks, dangling pointers and variable shadowing, and speeds up iterating through collections by \SIrange{20}{25}{\%}. Random access, which was slow with large linked lists, now completes in constant time. Depending on how often collections are iterated, typical workflows in \RooFit are sped up from \SIrange{5}{21}{\%}~\cite{ACAT19}. Fits with a binned ATLAS likelihood model~\cite{Hbb} completed \SI{19}{\%} faster while yielding identical results.
\bigskip

\noindent Modernising the C++ interfaces is also beneficial for using \RooFit from Python. Since \ROOT has a C++ interpreter, it can dynamically generate Python bindings for C++ objects~\cite{Pyroot}. In \ROOT 6.16, a loop over the parameters of a PDF would have to imitate the C++ code in the first listing. However, with an STL-like interface, Python iterators are generated automatically. The equivalent Python loop using \ROOT 6.18 therefore reads:
\begin{minted}{Python}
for p in pdf.getParameters(obs):
  p.Print()
\end{minted}
\bigskip

In \ROOT 6.22, the interfaces of category classes will be modernised in a similar way. Defining and printing category numbers and names compares as follows:

\begin{minipage}[t]{0.49\textwidth}
\centering \ROOT 6.22
\begin{minted}[fontsize=\footnotesize,highlightlines=5]{c++}
RooCategory cat("cat", "Lep. mult.");
cat.defineTypes(
    {"0Lep", "1Lep", "2Lep", "3Lep"},
    { 0,      1,      2,      3    });


for (const auto& name_idx : cat) {
  std::cout << name_idx.first << ", "
    << name_idx.second << std::endl;
}
\end{minted}
\end{minipage}
\hfil
\begin{minipage}[t]{0.49\textwidth}
\centering \ROOT 6.18
\begin{minted}[fontsize=\footnotesize]{c++}
RooCategory cat("cat", "Lep. mult.");
cat.defineType("0Lep", 0);
cat.defineType("1Lep", 1);
cat.defineType("2Lep", 2);
cat.defineType("3Lep", 3);

TIterator* typeIt = cat.typeIterator();
RooCatType* catType;
while ( (catType =
  dynamic_cast<RooCatType*>(typeIt->Next()) )
  != nullptr) {
  std::cout << catType.GetName() << ", "
    << catType.getVal() << std::endl;
}
delete typeIt;
\end{minted}
\end{minipage}

\subsection{Old Interfaces Remain Supported}
Given that the old interfaces of classes such as \texttt{RooAbsCollection} and \texttt{RooAbsCategory} (+ derived classes) are in use, they cannot be deprecated without forcing users to update existing code. Therefore, the new interfaces are provided as an addition, while old interfaces remain supported. Updating to the new interfaces leads to faster, shorter and more type-safe code, but it is not required.

To enable such kind of backward compatibility, legacy iterators/functions were implemented, which mimic the functionality of the original objects. Occasionally, this requires extra virtual calls or heap allocations, leading to slow downs of \SIrange{1}{3}{\%}, but the new interfaces allow for \SI{20}{\%} speed up. The most critical iterators in \RooFit have been modernised, and more iterators will be replaced as the modernisation of \RooFit continues. To detect uses of inefficient interfaces in user code, users can add \mintinline{c++}{#define R__SUGGEST_NEW_INTERFACE} before including \ROOT headers, which will trigger deprecation warnings with the \texttt{clang++}, \texttt{g++} and \texttt{MSVC} compilers. Further, old interfaces will be documented in special sections of \RooFit's reference guide~\cite{refGuide} to aid users in modernising existing code.

More updates of interfaces are planned, such as easier importing of data from \eg\ \ROOT's \texttt{RDataFrame}, STL containers or numpy arrays. The release of the new \texttt{PyROOT}~\cite{Pyroot} will further enable designing more pythonic interfaces that don't have to closely imitate the C++ syntax.

\section{More Stable and Faster Built-in \PDFs\label{sec:newPDFs}}
\begin{wrapfigure}[9]{r}{0.395\textwidth}
\vspace{-1\baselineskip}
\centering
\includegraphics[width=5cm,trim={8mm 0mm 0 14mm},clip=true]{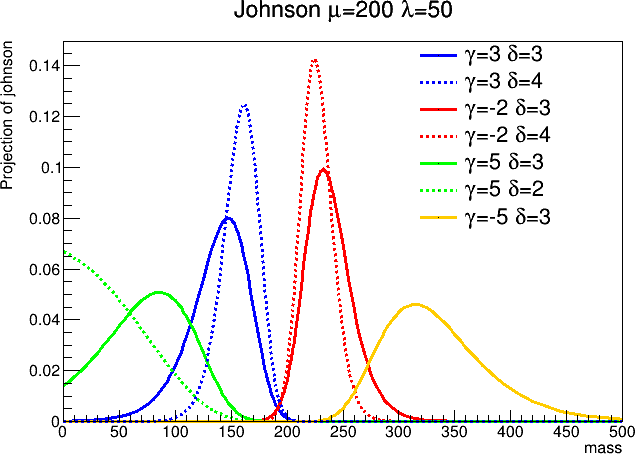}
\vspace{-1\baselineskip}
\hfil\caption{Johnson distribution}
\end{wrapfigure}
In modernising \RooFit, the Hypatia2~\cite{Hypatia} (\ROOT 6.20) and Johnson distributions~\cite{Johnson} (\ROOT 6.18, see figure) were added. Although any C++ function can be used as a \PDF in \RooFit (see \cref{sec:JITPDFs}), providing optimised implementations for these frequently used \PDFs is beneficial. Apart from providing a citeable default, built-in \PDFs can be extended with extra checks, and optimised computations can be applied. The advantages of built-in \PDFs are:
\medskip

\begin{description}
    \item[Analytic Integrals] In order to normalise functions, these must be integrated over the definition range of their observables. \RooFit integrates all functions numerically, unless a function for analytic integration is overridden. This allows for faster and more accurate normalisation.
    \item[Generating Samples] \RooFit generates data samples using the accept/reject method, unless a generator function is implemented. For many distributions, more efficient sampling strategies are known, which can only be employed for built-in \PDFs.
    \item[Checking parameters] Just-in-time compiled \PDFs as in \cref{sec:JITPDFs} will accept parameters and observables with arbitrary definition range. When trying to evaluate a function outside of its definition range, computations might yield negative probabilities, infinity or \texttt{NaN}. This can slow down or entirely prohibit fitting distributions to data, since the minimiser has no means of determining whether a parameter is in the allowed range. Starting with \ROOT 6.18, the definition ranges for parameters of built-in \PDFs can be checked by the \PDF implementation.
    \item[SIMD Computations] Starting with \ROOT 6.20, computations can be sped up significantly by evaluating large batches of data using \texttt{SIMD} computations. This is discussed in \cref{sec:fastPDFs}.
\end{description}
For the built-in Johnson distribution, for example, previously unstable fits were found to converge six times faster than with the commonly used interpreted Johnson formula in LHCb.

\section{Just-in-time Compiled \PDFs\label{sec:JITPDFs}}
Although implementing a \PDF as a \RooFit class is the fastest and most accurate way of building likelihood models, \RooFit was supporting interpreted \PDFs using the class \texttt{RooGenericPdf}. It takes strings of function expressions, and interprets these using an instance of \ROOT 5's \texttt{TFormula}. With the arrival of \texttt{cling}, the just-in-time C++ compiler and interpreter, \texttt{TFormula} was updated to use just-in-time compiled expressions. In \ROOT 6.20, the new \texttt{TFormula} was integrated in \RooFit, allowing for faster computations (compiled with optimisations instead of interpreted). This also enables using previously-defined functions or functions loaded from a library as in the following example:
\begin{minted}[highlightlines={7}]{c++}
// In a library or in included code
double func(double x, double a) { return a*x*x + 1.;}
[...]
// When building fit model
RooRealVar    x("x", "Observable", 2.); // Define observable
RooRealVar    a("a", "Parameter",  3.); // Define parameter
RooGenericPdf pdf("pdf", "func(x, a)", {x, a}); //evaluate func and normalise
\end{minted}

\section{Faster \PDF Computations\label{sec:fastPDFs}}
\subsection{Why Unbinned Fits Are Slow in \RooFit}
When \RooFit fits \PDFs to data, an expression is evaluated for each entry in the dataset to compute the likelihood of observing an event. 
\RooFit achieves this by writing values from the rows of a dataset into the leaves of a computation graph, and the probability of the top node is evaluated by calling evaluate and normalisation functions of daughter nodes. Each node caches its last value, and therefore constant branches of the computation graph (\eg\ branches that only 
depend on parameters) are only computed once.
This cycle, however, repeats for every entry in the dataset such that all branches that depend on observables have to be recomputed every time. The total number of function calls is therefore proportional to the number of entries in the dataset and
to the number of (non-constant) nodes in the graph, $N_\mathrm{Data} \cdot N_\mathrm{Nodes}$. For a small mathematical expression with 10 nodes and one million
events, this already amounts to considerably more than 10 million function calls because additional calls for the normalisation of PDFs and
for invalidating the node-local caches are necessary.

Furthermore, loading only single values into the nodes of the computation graph is hostile to CPU caches. It is possible that
when a node is being revisited to load the next entry, data have been evicted from the cache(s). This means that \RooFit runs inefficient on
modern CPUs because of poor data locality and inefficient memory access patterns.

\subsection{Three Times Faster Computations with Higher Data Locality}
To improve the data locality and reduce the number of function calls, a \RooFit-internal interface for batched likelihood computations was implemented. Data are passed between nodes using a \texttt{std::span}. Inputs for computations are directly read from arrays, while outputs are written to contiguous memory that is owned by the node that is running a computation.

\begin{figure}[t]
\begin{center}
\includegraphics[width=1.\linewidth,trim={0 3mm 0 10mm},clip=true]{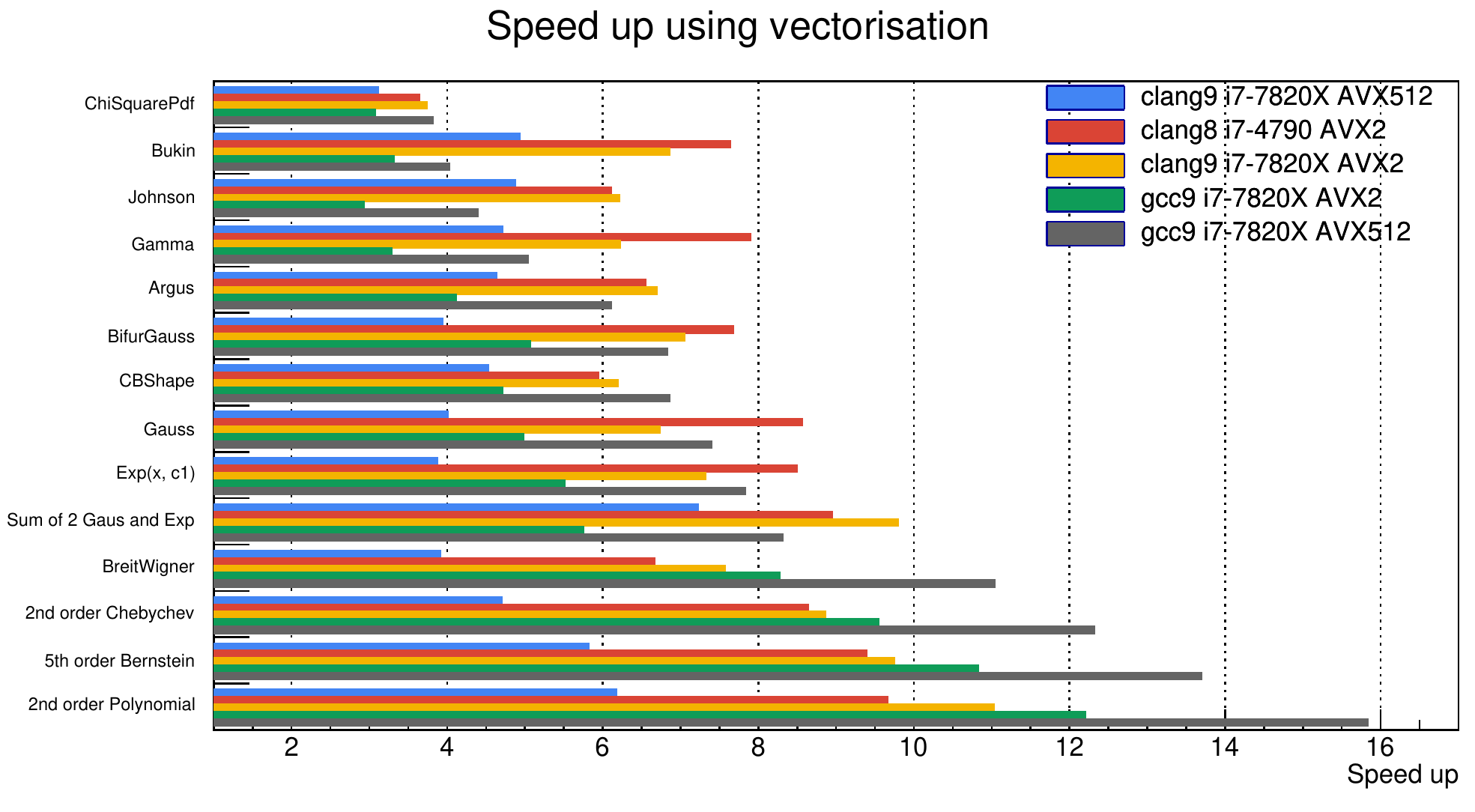}
\end{center}
\caption{Speed up for computing the likelihoods of datasets of $100\,000$ to $300\,000$ events for various likelihood models. Using \ROOT 6.20, the fast \RooFit batch interface is timed against the normal single-value computations for different CPUs, compilers and instruction sets. On an intermediate-level CPU that supports \texttt{AVX2} instructions ({\setlength{\fboxsep}{0pt}\framebox{\textcolor[RGB]{219,68,53}{\rule{1em}{1ex}}}}), a speed up of 4x to 9x can be expected. On CPUs supporting \texttt{AVX512} instruction sets, the speed up ranges from 4x to 16x. For the ``ChiSquarePdf'' (top), almost no \texttt{SIMD} functions were available. The speed up of 3x is the result of batched computations with faster data loading.}
\label{fig:vecSpeed}
\end{figure}

Instead of computing only one probability per node, all probabilities for all entries in the dataset can be computed in a few function calls.\footnote{\RooFit also supports a multi-process mode, in which case smaller batches of contiguous data are processed, which are divided among multiple workers.} Since such computations operate on contiguous floating point numbers, data locality, caching and data prefetching improve. This speeds up computations three to four times without loss of precision. \Cref{fig:vecSpeed} shows the combined speed up of batched computations and additionally vectorisation with \texttt{SIMD} instructions (\cref{sec:SIMD}) against classic single-value \RooFit computations. For the topmost entry (``ChiSquarePdf''), almost no \texttt{SIMD} instructions could be used. This speed up is mostly due to faster data loading and reducing function calls.

When \PDFs that support the fast interface are used together with legacy \PDFs, compatibility is ensured by
a generic batch computation function that runs \RooFit's classic single-value computations for each entry in the dataset, writes results into an array, and passes the results on to the next node of the graph using \texttt{std::span}.

\subsection{4x to 16x Faster Computations with \texttt{SIMD}\label{sec:SIMD}}
By converting \RooFit's data access patterns to reading from array-like structures, \RooFit could be extended with single-instruction-multiple-data computations (\texttt{SIMD}). Optimisation reports of the \texttt{clang++-8} compiler were used to design computation functions that allow for automatic vectorisation with \texttt{AVX2} instructions. This allows to increase the throughput, since four double-precision numbers can be processed simultaneously (8 with \texttt{AVX512}).
This requires using auto-vectorisable, inlinable mathematical functions, and entries in arrays 
cannot depend on other entries. Auto-vectorisable functions are provided by the
\texttt{VDT}~\cite{VDT} package, of which logarithm and exponential function were used most frequently. \texttt{VDT} approximates standard library functions using Pad\'e polynomials, which can be inlined into other computations, and optimised by the compiler. \ROOT can be compiled without \texttt{VDT}, though, in which case the standard (non-inlinable) implementations are used. This usually prevents automatic vectorisation.

To further aid the compiler optimiser, computations were rewritten such that data dependencies between elements are eliminated, and that as many intermediate results as possible can be held in CPU registers. Excessive branching in complicated functions was replaced by branch-less computations such as $1 \cdot A + 0\cdot B$ to select $A$ and its counterpart to select $B$. Although more CPU cycles are needed because both branches are computed, higher throughput is achieved because \texttt{SIMD} instructions process four or eight entries simultaneously. Finally, calls to non-inlinable functions and complicated reductions were removed where possible.

The result of this work is shown in \cref{fig:vecSpeed}. For various \RooFit \PDFs and composite models that use multiple \PDFs, the run time of the optimised \texttt{SIMD} computation with \texttt{VDT} functions is compared to classic \RooFit single-value computations. Depending on the complexity of the computations and on the level of automatic compiler optimisations, computations speed up 3x to 16x. The theoretical speed up of 12x (3x from data loading, 4x for \texttt{AVX2}) cannot be reached because of Amdahl's law. Not all parts of \RooFit's code can be replaced with \texttt{SIMD} code, and were therefore not sped up. Moreover, in order to use \texttt{SIMD} instructions, more instructions may have to be issued (\eg\ compute two Pad\'e polynomials to approximate the \texttt{exp} function instead of calling the standard implementation).

The level of automatic compiler optimisations has a strong effect on the speed up. \Cref{fig:vecSpeed} demonstrates that both \texttt{clang} 8 and 9 have powerful optimisers for \texttt{AVX2} instructions, but \texttt{gcc} significantly outperforms \texttt{clang} when \ROOT is compiled with \texttt{AVX512} instructions. It can therefore be expected that as compiler optimisers improve, the speed up for \RooFit increases, and that differences between the compilers will shrink.
\bigskip

Batch computations with vectorisation have been released in \ROOT 6.20. These can be switched on by adding the ``BatchMode'' argument to the fitting instruction:
\begin{minted}[escapeinside=||]{c++}
pdf.fitTo(data, |\textcolor{blue}{BatchMode}|(true)); // Evaluate likelihood using fast batch mode
\end{minted}
Unit tests ensure that the optimised computation functions yield the same results as the classic \RooFit functions. When the fast \texttt{VDT} approximations are used, the relative difference for probabilities is usually below \num{1.E-14}, for log-likelihoods below \num{2.E-14}, and for fit parameters below \num{1.E-5}, which is except for corner cases orders of magnitude smaller than the statistical error of the fit parameters.

A drawback of the current implementation is that for maximal performance, \ROOT has to be compiled targeting the instruction set that the CPU supports. The possibility of shipping a small library with computation functions for a few different architectures (\eg\ \texttt{SSE4} + \texttt{AVX2}) is being investigated.

\section{Summary}
\RooFit's interfaces are being modernised, and computations are being sped up with fast data loading and \texttt{SIMD} computations. The single-thread performance of unbinned fits in \RooFit has been increased by several factors without requiring code changes on the user side.
In conjunction with work on parallelising computations~\cite{Patrick}, a speed up by more than an order of magnitude can be expected.

The work on modernising \RooFit's interfaces and improving its performance will continue, especially for binned fits such as \HistFactory models.

\newpage
\bibliography{CHEP19}

\end{document}